# LORENTZ CONTRACTION ACCORDING TO THE WAVE NATURE OF THE LIGHT

Giovanni Zanella

*Dipartimento di Fisica dell'Università di Padova and Istituto Nazionale di Fisica Nucleare, Sezione di Padova, via Marzolo 8, 35131 Padova, Italy*

**Abstract**
*The wave nature of the light, applied to the kinematics of the moving bodies, permits to investigate and find a coherent solution, on some questions raised by the theory of special relativity about the Lorentz contraction.*

## 1. Introduction

The so-called Fitzgerald-Lorentz contraction (simply *Lorentz contraction*) as presented by Einstein in his special theory of relativity [1][2] has been always object of discussion. See, for example, the recent papers of S.D. Agashe [3], R.D. Klauber [4] or of Y. Pierseaux [5] and their references, where the possibility of a dilation instead of a contraction is not excluded. On the other hand the Lorentz contraction has never been directly observed. Analyzing the theory of the special relativity (SRT) some in-depths appear necessary about this issue.
We shall found our analysis on the following hypotheses:
 1. *The empty space is isotropic.*
 2. *The empty space is not a "stationary" support of physical events and a "stationary" reference system does not exist.*
 3. *The physical events occurring in the same spatial point and in the same instant (coincident events) are inseparable if viewed from any other reference system.*
 4. *The physical events occurring in a system of reference are not affected by the existence of any other non-interacting system of reference.*
 5. *The path of the light records the time of the system where the light is emitted (measured locally or from another system).*

These hypotheses track the two known postulates of SRT, or they are there tacitly intended. Insuperability and constancy of the velocity of the light, no matter the reference system, appears to be just a consequence of such hypotheses [5].

Another consequence, from these hypotheses, is that the motion of a body can never be intended independently from the symmetrical and opposite motion of a body of reference (*principle of reciprocity* or *of symmetry*).

In the following we shall use Cartesian systems of coordinates (supposed embedded in material bodies) and the methods of Euclidean geometry, where the terms: *view*, *see*, *observe*, *appear*, etc. have qualitative meaning, while the result of a *measurement* is quantitative.

Suppose now to have two Cartesian systems (S and $S_0$) with their axes in parallel, being coincident the x-axes. Suppose $x$ be the positive coordinate of one end of a rigid rod, lying in parallel to x-axis of S (Fig.1) while the other end lies in the origin of the system. Suppose also to have another rigid rod, of the same length, lying on the positive x-axis of $S_0$, with one end put in its origin.

Now, impart a uniform motion of parallel translation to the system S, in respect to $S_0$ and in the direction of increasing x, letting a spherical wave of light be emitted from the origins of S and $S_0$ when they are coincident. Hence, *an observer at rest in $S_0$ and an observer at rest in S see either the same spherical wave front to reach the end of the rods in coincidence* (Fig.1). In other words, an observer in S sees the wave front to reach the end of own rod exactly when an observer in $S_0$ sees the same wave front to reach the end of the own rod. Indeed, the systems are perfectly equivalent in respect to the motion of the spherical wave front and it is not possible to privilege a system in comparison to the other.

As a consequence, the rod fixed in S would appear contracted to the observer in $S_0$ (Fig.1) and the rod fixed in $S_0$ would appear dilated to the observer of S (Fig.2). But, this result cannot be accepted, because it lacks of symmetry. Indeed, the position of the origin of the axes of S and $S_0$ on the rods is arbitrary and it requires, for reasons of symmetry, to be put in the middle of the rods. Therefore, as we shall see, the final result will be a contraction accompanied by a dilation, so that the whole length of the moving rod will be dilated by a scale factor (the *Lorentz factor*).

Einstein proposed in his paper [1] another method to measure the length of a moving rigid rod: *stationary clocks are supposed synchronized in the stationary system* (system $S_0$)*, then an observer ascertains at what points of the stationary system the two ends of the moving rod* (system S) *are located* *at a definite time. The distance between these two points, measured by the ruler of stationary system, is the length of the moving rod.*

Now this measurement supposes the possibility of synchronized clocks indifferent spatial points and of a stationary system. Instead, we will show the physical impossibility to reach the synchronization of clocks put in different spatial points.

About the shape of a moving rigid sphere, this remains invariant. Indeed, if we look to a rigid sphere at rest in the moving system S, with its centre in the origin of the coordinates and let a spherical wave be emitted from the origins of S and $S_0$ when they are coincident, *an observer at rest in $S_0$, and an observer at rest in S, will see both the same spherical wave front to reach the surface of the sphere in coincidence.* In other words, this wave front will reach only once the surface of the sphere and such event is the same for both the observers. So the shape of a moving rigid sphere will appear invariant to the two observer.

As concerns the diameter of the sphere, or the length of the rod, when they are in motion, their sizes will appear dilated by the Lorentz factor, if they are cross-measured from a system to the other.

The deduction of Einstein [1] about the contraction of the diameter of a moving rigid sphere, which center lies in the origin of the system S, where it is at rest, suffers of the same problem of the measurement of the length of a moving rod, because he assumes that the ray of the sphere is measured when the center of the sphere coincides with the origin of the system $S_0$, that is at the same time *t*=0.

At least, about the time, being it intended as the path of the light measured locally in the system of the source or from an other reference system, it follows the fate of the dilation of a moving rod. But, it is necessary to note that the measurement of the time is not just as the measurement of the size of a rod, because the time is a cumulated quantity until the present, starting from a precise instant in the past. Instead, a length of a rod is measured always at the present, without involve its elapsed history.

**2. Clock synchronization**

Einstein in his paper proposed to synchronize two equal clocks put in different points A and B of space using the following procedure. *Let a ray of light start at the "A time" $t_A$ from A towards B, let it at the "B time" $t_B$ be reflected at B in the direction of A, and arrive again at A at the "A time" $t''_A$.*

*Therefore the two clocks synchronize if*

$$t_B - t_A = t''_A - t_B \quad . \tag{1}$$

Relationship (1) includes the necessity that the two clocks march according the path of the light, in the sense that $t_B - t_A$ is proportional to the light path from A to B, and vice versa. Indeed, the double path of the light (forward and back) corresponds to a double increment of the time, that is

$$t''_A - t_A = 2(t_B - t_A) \quad . \tag{2}$$

Therefore

$$t_B - t_A = \frac{AB}{c} \quad , \tag{3}$$

where $c$ is velocity of the light.

Einstein's procedure permits to test only the rate of two clocks put in different places of the space, but it does not test their starting points. Indeed, Eq.(1) is invariant changing the starting point of $t_A$ and of $t_B$.

Therefore, we demand us if it is possible to synchronize clock A with clock B in manner that $t_A = t_B$. So, excluding a synchronization of the clocks in A followed by the displacement of one clock from A to B (because the synchronization is not guaranteed after the trip), we suppose to adjust clock B according to clock A, following this procedure. Clock B can at best be informed of the time $t_A$ with a signal of velocity $c$. Hence, sending a ray of light from A to B when $t_A = 0$, this ray will reach B at the time $t'_A = AB/c$ and if $t_B$ is not equal to $t'_A$ it is necessary to restart clock B. The restarting of clock B requires to put $t_B = 0$ just when it is reached by the light from A, because clock B cannot be restarted, and synchronized with clock A, at $t_A = 0$.

A part the problematic quantification of $AB/c$, which would require knowledge of the distance AB, the picture the clocks put in different points of the space, which measure all the same time is not physically consistent also under another point of view. Indeed, it is known that LT transform any set of four space-time co-ordinates, maintaining the invariance of the *space-time interval* between two events. The intervals pertaining two events connected by a hypothetical signal travelling at velocity $> c$ are named *like-space*, while the others are named *like-time*. Obviously, for two events connected by a *like-space* interval it is physically impossible to define their temporal order, as well as their simultaneity.

In conclusion, it is not correct to define the length of a rigid rod measuring the distance of two events, put in different places of the space, *at the same time*, that is connecting them with a signal of infinite velocity.

In the following, we shall use *light-clocks*, where the time is measured by the path of the light along a graduated line, a part the constant of proportionality $1/c$ [5]. The use of light clocks, as we shall see, permits us to discover that a moving light clock furnishes results which depend on the orientation of the wave of light in respect the direction of its motion.

### 3. The factor of Lorentz

In the analysis of the Lorentz contraction, the *Lorentz factor* plays a fundamental role. Therefore, it is useful to deep its meaning. In Fig.1, the distance $x$ of the spherical wave front propagating along the positive x-axis of S and $S_0$ (supposed "stationary"), is

$$x = x_o - ut_o \ , \tag{4}$$

where $u$ is the velocity of S in respect to $S_0$ while $x_0$, $t_0$ are referred to $S_0$.

In realty, the system $S_0$ is not "stationary" because, symmetrically, we can consider equivalent the system $S_0$ moving with velocity $-u$ and "stationary" S (Fig.2). Therefore, the result of the measurement of $x$ from $S_0$ will require a factor of correction $\gamma$ (*Lorentz factor*), that is

$$x = \gamma (x_o - ut_o) \ , \tag{5}$$

where $\gamma$ has to be determined according to the *principle of reciprocity* (the motion of S with velocity $u$, in respect to S, is equivalent to the motion of $S_0$ with velocity $-u$ in respect to S. In other words, an observer at rest in $S_0$, in motion with velocity $-u$, see S exactly as the observer at rest in $S_0$ sees S moving at velocity $u$).

Besides, at the same manner of the coordinate $x$, the time $t$ of S, observed from $S_0$, is (see the path of the light in Fig.1)

$$t = \gamma \left( t_o - \frac{ut_o}{c} \right) = \gamma \left( t_o - \frac{ux_o}{c^2} \right) \ , \tag{6}$$

where the last term of Eq.(6) is obtained putting $t_o = x_o/c$, being $x_o$ the co-ordinate of the front of the light measured from $S_o$. Obviously, the *Lorentz factor* of the space is used also for the time, being spatial the scale of the light clock.

At this point, we can multiply both the sides of Eq.(6) with $u$ and sum Eq.(5) with Eq.(6). Thus, we have $x + ut = \gamma x_o (1 - u^2/c^2)$. So

$$x_o = \frac{x + ut}{\gamma\left(1 - \dfrac{u^2}{c^2}\right)} \quad , \tag{7}$$

but, for the *principle of reciprocity*, we can calculate $x_o$ symmetrically (Fig.2)

$$x_o = \gamma(x + ut) \quad , \tag{8}$$

where $x$ and $t$ are measured from $S_0$.

Therefore, comparing Eq.(7) with Eq.(8), we obtain the *Lorentz factor*

$$\gamma = \frac{1}{\sqrt{1 - \dfrac{u^2}{c^2}}} \quad . \tag{9}$$

Vice versa, from Eq.(8), putting $t = x/c$ and $x = \gamma x_0 \left(1 - \dfrac{u}{c}\right)$, we find again $x_0$ of Eq.(7).

## 4. Distortion of a moving rigid body

If the shape of a moving rigid sphere appears invariant, the same thing cannot be affirmed for the size of its diameter when it is measured from the reference system. Indeed, consider a rigid rod (long as the diameter of the sphere and virtually embedded in it) put in parallel to the x-axis of the moving system S, where the origin of S lies in the middle of the rod. If we now impart a uniform velocity $u$ to the rod in the direction of increasing x, the half of the moving rod placed in the positive x-axis of S will result contracted (Fig.3 and Fig.4), that is

$$x = \gamma(x_0 - ut_{02}) = \gamma x_0 \left(1 - \frac{u}{c}\right) = x_0 \frac{\sqrt{1 - \frac{u}{c}}}{\sqrt{1 + \frac{u}{c}}}. \tag{10}$$

On the other hand, the half of the rigid rod oriented in the direction of the decreasing x-axis of S will appear dilated (Fig.1) according to

$$d = \gamma x_0 \left(1 + \frac{u}{c}\right) = x_0 \frac{\sqrt{1 + \frac{u}{c}}}{\sqrt{1 - \frac{u}{c}}}. \tag{11}$$

In conclusion, the whole length $<x>$ of the rod will be dilated. Indeed

$$\langle x \rangle = \gamma \frac{x_0}{2}\left(1 - \frac{u}{c}\right) + \gamma \frac{x_0}{2}\left(1 + \frac{u}{c}\right) = \gamma x_0, \tag{12}$$

where $\gamma$ is the Lorentz factor which assumes the meaning of a scale factor (Fig.4). Symmetrically, a rod put in $S_0$ and moving with velocity $-u$ will be also dilated, if viewed from S.
The distortion of a rigid system of rods put in a plane and in various directions, fixed in the origin of S, moving with velocity $u$ in respect to $S_0$ (Fig.5) will be as in Fig.6, if viewed from the system $S_0$, a part the factor of scale $\gamma$. Obviously, such distortion will be mirror in the symmetrical condition.

## 6. Time dilation

The dilation of a moving rigid rod is not a distinct phenomenon from the flow of the time, if the time is the path of the light out coming from a source. Looking to Fig.1, two opposite one-dimensional paths of the light appear, starting from the origin of the moving system. Therefore, the mean time of S, as measured from $S_0$, which symmetrically moves in the opposite direction, will be (Fig.2)

$$\langle t \rangle = \frac{1}{2}\left[\gamma t_0(1-\frac{u}{c}) + \gamma t_0(1+\frac{u}{c})\right] = \gamma t_0 \quad , \tag{13}$$

being $t_0$ the local time signed by the clock of S.
Now, we can use mechanical clocks to measure <t> simulating with them the average behavior of the light clocks.
As a consequence, the time elapsed during one hour (as signed on the dial of the clock by its hands) is elongated beyond one hour if measured from the system of reference in respect of which the clock is moving. In particular, the time is dilated until to become infinite if the velocity is $c$.
 Therefore, the time signed by a moving clock B, on its dial, is in delay in respect to the time signed on the dial of the clock considered as the reference system (clock A).
 At least, if clock B moves along a polygonal line and it returns to the starting point, at rest with clock A, clock A will measure <t> while clock B will measure $t_0$. In particular, if $t_0 = t_{01} + t_{02} + \cdots + t_{0n}$ and during these intervals the velocity of clock B is $u_1, u_2, \cdots, u_n$, the corresponding factors of Lorentz will be $\gamma_1, \gamma_2, \cdots, \gamma_n$, so

$$\langle t \rangle = \sum_{1}^{n} {}_i \gamma_i\, t_{0i} \quad . \tag{14}$$

Symmetrically, if clock A (system $S_0$) moves with velocity $-u$, in respect to clock B (system S), now <t> will represent the time of B and $t_0$ the time of A, so that clock A is delayed compared with clock B. Indeed, symmetry requires that the two clocks run in opposite direction to the previous. Vice versa, if the inversion of the march the two clocks it is not possible, the symmetry cannot be verified.

## 7. Conclusions

Since, all moving bodies are potential light sources, their motion not only deforms their inside and the external shape, but also the spatial scale and the space on the outside of them. So, the wave picture of the light permits to deepen the question of the Lorentz contraction, arriving to a coherent result in accord with the principle of symmetry. Besides, as the time corresponds to a measurement of the spatial path of the light, its dilation is not different from the dilation of a rigid rod moving in the same direction. At least, it appears interesting to verify as, following this picture, the known paradoxes of SRT find their solution.

## Figure captions

**Fig.1** Wave fronts of a pulse of light emitted from a source at rest in the moving system S and started from the origins of S and $S_0$ when they are coincident, viewed by an observer at rest in $S_0$.

**Fig.2** Wave fronts of Fig.1 viewed by an observer at rest in S.

**Fig.3** Coefficients of contraction.

**Fig.4** Symmetrical dilation of a moving rigid rod: (a) dilation of a rod at rest in S moving with velocity $u$ in respect to $S_0$ (as viewed from $S_0$); (b) dilation of the same rod at rest in $S_0$ moving with velocity $-u$ in respect to S (as viewed from S).

**Fig.5** Rod array viewed locally.

**Fig.6** Rod array of Fig.5 (at rest in S) in motion with velocity $u$, as viewed from the system $S_o$.

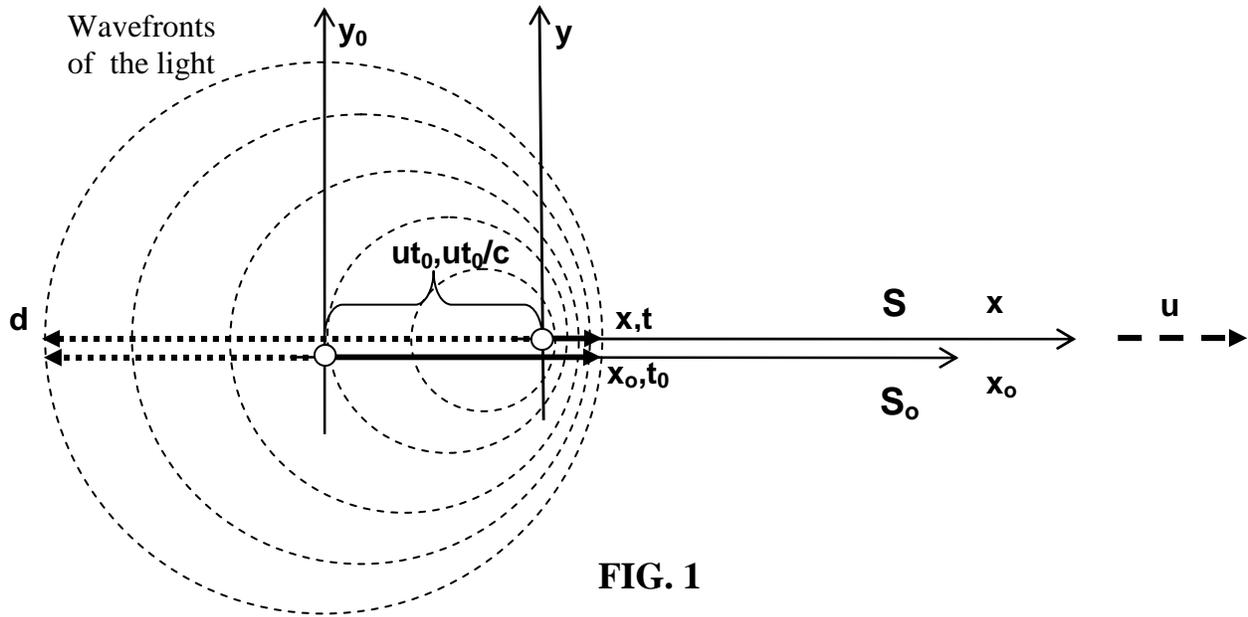

**FIG. 1**

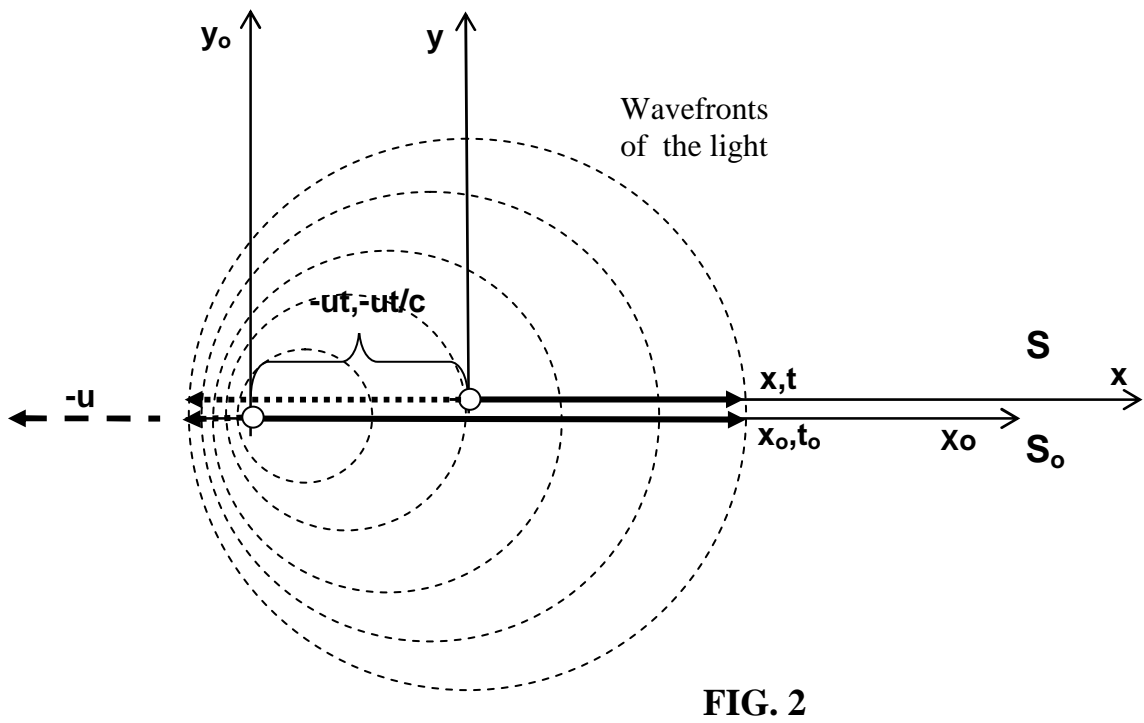

**FIG. 2**

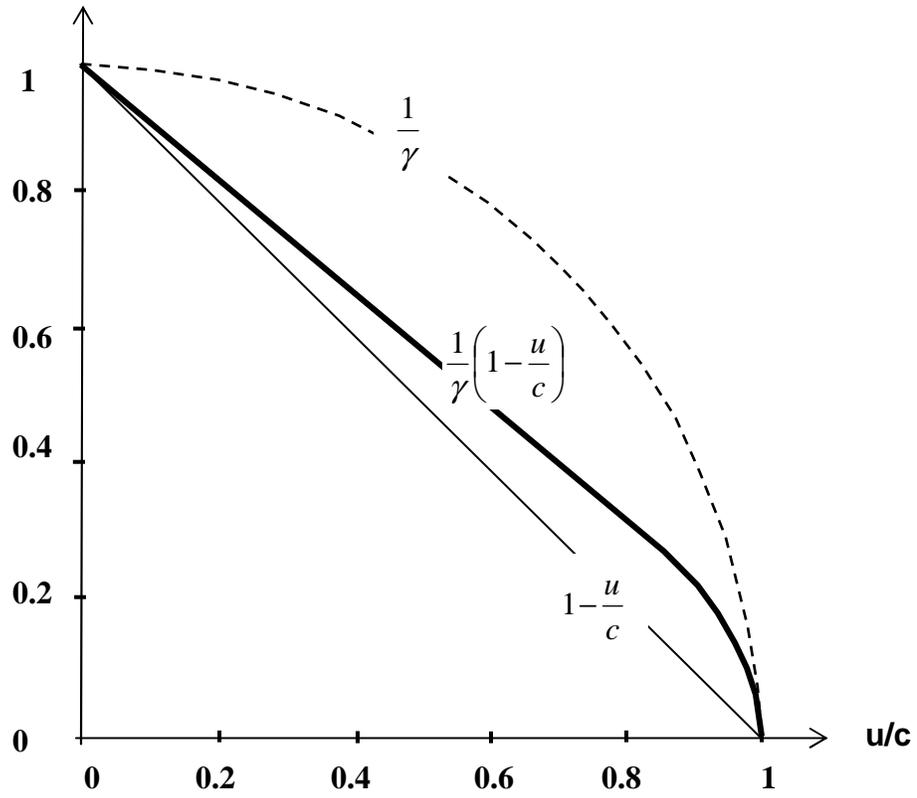

**FIG. 3**

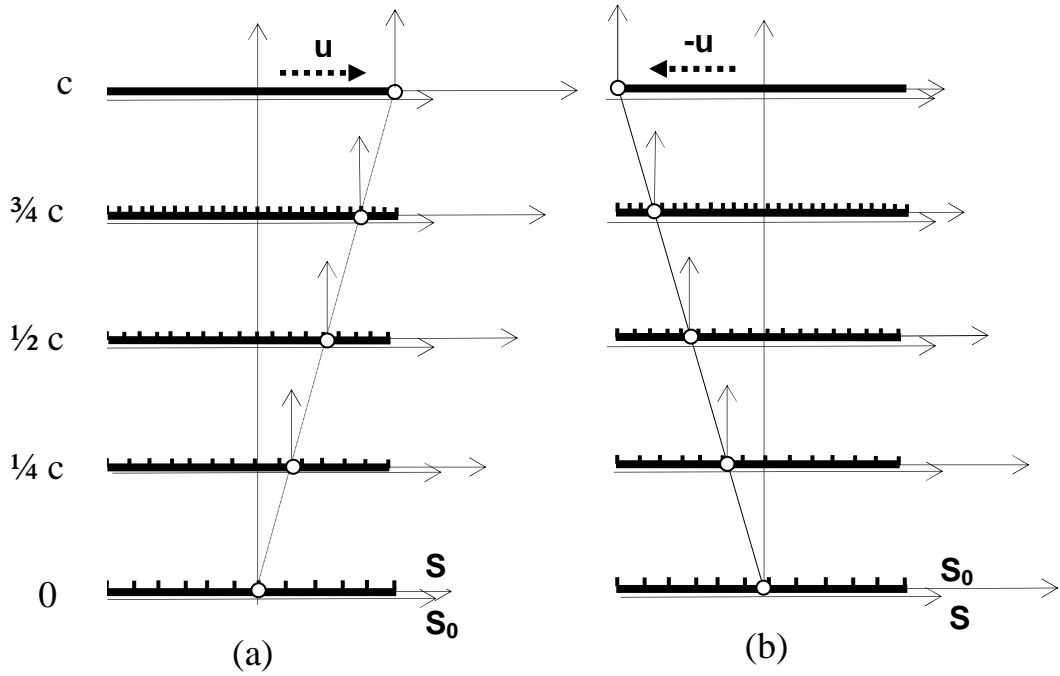

**FIG. 4**

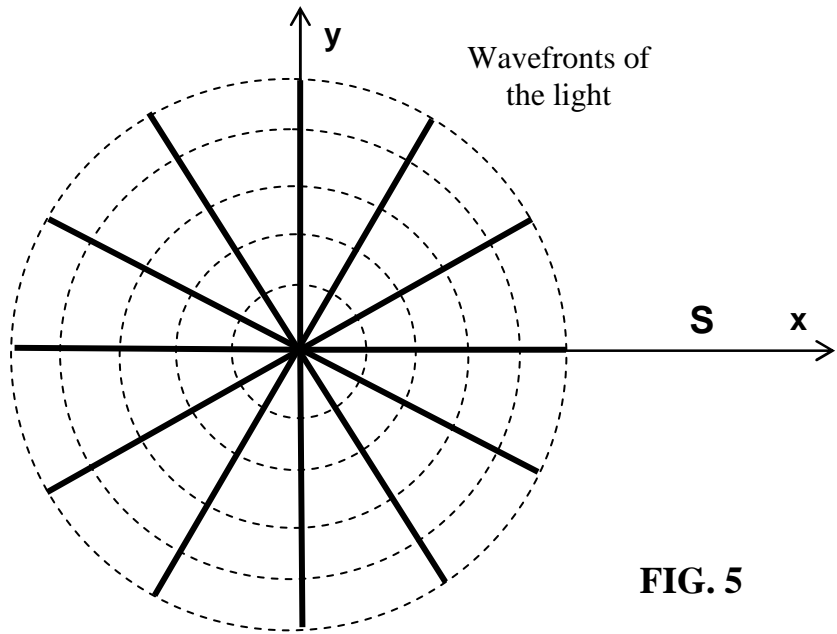

**FIG. 5**

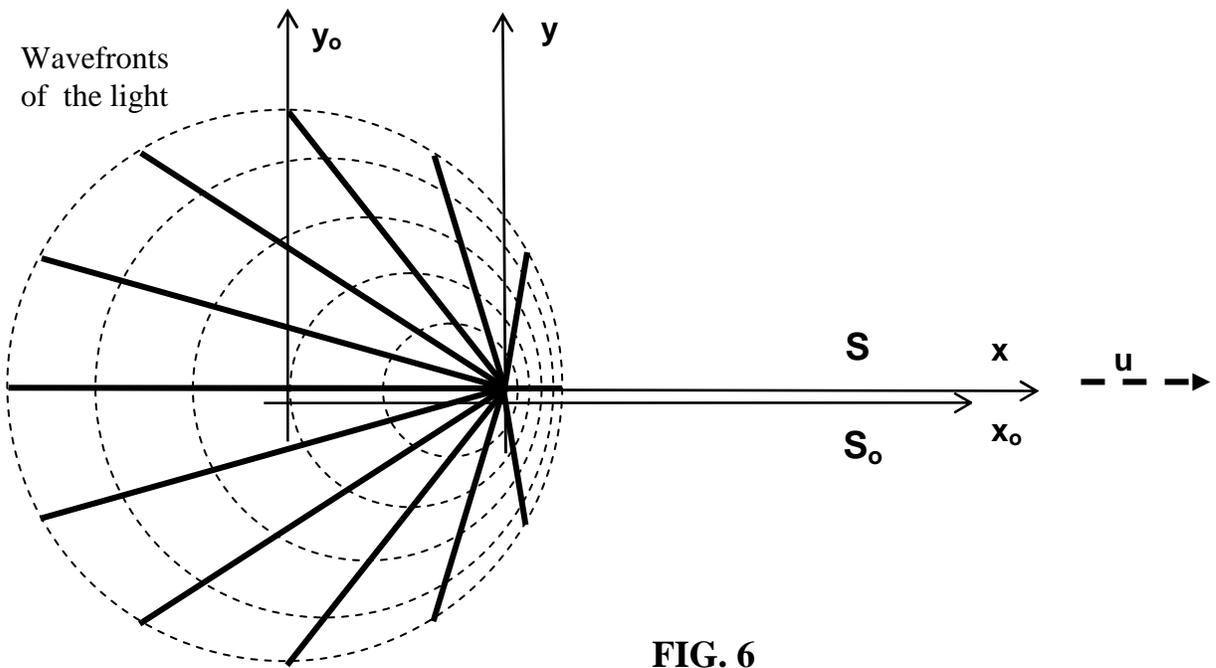

**FIG. 6**